\documentclass[10pt,conference]{IEEEtran}
\IEEEoverridecommandlockouts
\usepackage{amsmath,amssymb,amsfonts}
\usepackage{enumitem}
\usepackage{algorithmic}
\usepackage{graphicx}
\usepackage{textcomp}
\usepackage{xcolor}
\usepackage{xspace}
\usepackage{multirow,tabularx}
\usepackage{ragged2e}
\usepackage{threeparttable}
\usepackage[hyphens]{url}
\usepackage{balance}
\usepackage[sort, square, numbers]{natbib}
\usepackage[many]{tcolorbox}
\tcbuselibrary{theorems}
\definecolor{custom-gray}{cmyk}{0, 0, 0, 0.7, 1.00}
\newtcbtheorem[no counter]{Summary}{\hskip-0.97em}{enhanced,drop shadow={black!50!white},
 coltitle=white,
 top=0.15in,
 bottom=0.05in,
 attach boxed title to top left=
 {xshift=0.5em,yshift=-\tcboxedtitleheight/2},
 boxed title style={size=small,colback=custom-gray},
 separator sign none,
 boxsep=0pt
}{summary}

\newcolumntype{Y}{>{\centering\arraybackslash}X}
\def\BibTeX{{\rm B\kern-.05em{\sc i\kern-.025em b}\kern-.08em
    T\kern-.1667em\lower.7ex\hbox{E}\kern-.125emX}}

\newcommand\xapp{FMware\xspace}
\newcommand\xapps{FMware\xspace}
\newcommand\xjudge{AI judge\xspace}
\newcommand\xjudges{AI judges\xspace}
\newcommand\xjudgesystem{\xjudge system\xspace}
\newcommand\xjudgesystems{\xjudge systems\xspace}

\begin{document}

% \title{Engineering AI Judge Systems: Life Cycle, Challenges and Framework}
 \title{Engineering AI Judge Systems}

\author{\IEEEauthorblockN{Jiahuei (Justina) Lin, Dayi Lin, Sky Zhang}
\IEEEauthorblockA{\textit{Centre for Software Excellence, Huawei Canada} \\
Kingston, Canada \\ 
cse@huawei.com}
% \and
% \IEEEauthorblockN{Dayi Lin}
% \IEEEauthorblockA{\textit{CSE, Huawei} \\
% Kingston, Canada}
\and
\IEEEauthorblockN{Ahmed E. Hassan}
\IEEEauthorblockA{\textit{Queen's University} \\
Kingston, Canada\\
ahmed@cs.queensu.ca}
}

\maketitle

\begin{abstract}
% Recent emergence of building foundation model-based (FM-based) applications 

\xjudgesystems are designed to automatically evaluate Foundation Model-powered software (i.e., \xapp). Due to the intrinsic dynamic and stochastic nature of \xapp, the development of \xjudgesystems requires a unique engineering life cycle and presents new challenges. In this paper, we discuss the challenges based on our industrial experiences in developing \xjudgesystems for \xapp. These challenges lead to substantial time consumption, cost and inaccurate judgments. We propose a framework that tackles the challenges with the goal of improving the productivity of developing high-quality \xjudgesystems. 
Finally, we evaluate our framework with a case study on judging a commit message generation FMware. %, that is important for program comprehension and the maintenance of a software project. 
The accuracy of the judgments made by the \xjudgesystem developed with our framework outperforms those made by the \xjudgesystem that is developed without our framework by up to 6.2\%, with a significant reduction in development effort.
\end{abstract}


\section{Introduction}
\label{sec:intro}
The rapid rise of Foundation Model (FM) powered software (i.e., \xapp) has led to a growing need for robust evaluation mechanisms. However, given the open-ended nature of these responses of FMs that are part of an \xapps, it is difficult for developers to either manually evaluate all the responses, or craft a test dataset that includes all the possible responses of an \xapp as oracles, and use the test dataset to measure the quality of the \xapps. Therefore, practitioners have been developing FM-based evaluators (i.e., \xjudges) to reduce the manual evaluation effort for these \xapps. 

% [rephase from the dev perspective]
% [AI-as-a-judge pros]
Developing \xjudgesystems for the automatic evaluation of \xapps offers several advantages. First, \xjudges have shown to be capable of evaluating open-ended responses and can generate detailed explanations for their judgments~\cite{zheng2023judging, dubois2024alpacafarm, dettmers2024qlora}. As the FMs in \xjudges were trained on vast corpora, these models possess expertise in natural language understanding and instruction following, allowing them to measure various aspects of text quality effectively. Second, compared to human evaluators, \xjudges require significantly less time for evaluations, enabling the evaluation of a large amount of data quickly~\cite{bubeck2023sparks, zheng2023judging}. Additionally, \xjudges are not affected by factors that are associated with human annotators (e.g., fatigue, emotions, and distractions). 

However, translating high-level requirements to measurable units is inherently challenging. Unlike conventional software applications, where judging requirements are often directly tied to specific functionalities or outputs, \xjudgesystems for \xapp must grapple with abstract concepts such as fairness, accuracy, and contextual understanding. These high-level requirements can be difficult to operationalize, leading to ambiguity in how an \xapp should be evaluated. As a result, developers must engage in a complex iterative process to refine these requirements into quantifiable metrics~\cite{pan2024human}.  This involves not only extensive testing and calibration of the judging systems \cite{zheng2023judging, liu2023calibrating} but also a deep understanding of the underlying \xapp that are intended to be assessed. The labor-intensive nature of this task can slow down the development process and impact the overall effectiveness of the \xjudgesystems.

In addition, developers must navigate the intricacies of developing \xjudgesystems through a combination of cognitive architectures for evaluation and various jury FMs. We refer to judging cognitive architecture that includes components such as jury FMs, judging heuristics, prompts and their relations, as well as metrics and their interactions, to make accurate judgments. There is no universally applicable judging architecture for all \xapp. Each target \xapp may require unique judging cognitive architectures tailored to its specific characteristics and objectives. This requires a flexible design that can incorporate diverse cognitive architectures, which may vary significantly in their methodologies and underlying assumptions. Furthermore, as the \xapp field continues to evolve rapidly, \xjudgesystems must be adaptive, continuously integrating new judging techniques into their judging architectures to keep pace with these changes. Such a dynamic environment often results in inconsistent judgments, particularly when a target \xapp undergoes updates. As such, developers must maintain a balance between establishing robust judging architectures and ensuring that their \xjudgesystems remain accurate to ongoing developments in the \xapp landscape.

In this paper, %we discuss the emergence of constructing \xjudgesystems to ensure the quality of \xapps. 
% we first introduce the life cycle of \xjudgesystems. %and their associated engineering stages for the construction of \xjudgesystems. 
% For each stage, 
we discuss the challenges that developers face when defining judging requirements, developing an \xjudgesystems, and evolving it. We highlight the difficulty of applying mitigations for these challenges based on insights gained from the in-depth discussions with academic and industrial leaders. These insights were gathered from events including (1) \textit{SEMLA 2023} and\textit{ 2024}~\cite{semla2023}, (2) \textit{FM+SE Vision 2030}~\cite{fmse2023vision}, (3) \textit{SE 2030 workshop--FSE 2024}~\cite{se2030workshop, fse2024}, (4) \textit{FM+SE Summit 2024}~\cite{fmse2024summit} and (5) through our active contributions to the Open Platform for Enterprise AI (OPEA) initiative~\cite{opeaOpenPlatform}. OPEA is a multi-organization collaborative effort, including industry leaders like Intel, AMD, Docker, ZTE, RedHat, SAP, Anyscale, Llamaindex, and others, and focuses on addressing the challenges around making generative AI-powered software enterprise-ready.

% These challenges were identified based on (i) surveys of academic and grey literature, (ii) in-depth discussions with industrial and academic leaders (e.g., during SEMLA 2023~\citep{semla2023} and the FM+SE Vision 2030~\citep{fmse2023vision} event with over 100 attendees from many leading companies), (iii) meetings with our customers and our development teams to understand the functional and non-functional needs of their \xware, and (iv) our practical experience in designing, implementing, and maintaining our in-house \xware lifecycle engineering platform (\fmarts) and several complex \xware for strategic customers.

We then propose a search-driven constitution-based framework for developing \xjudgesystems. The framework abstracts the required elements and steps to address the challenges. We then evaluate the value of the framework through a case study on judging an \xapp that generates commit messages.
% We propose a constitution-driven framework for constructing \xjudgesystems designed to ensure the quality of judgements. 
The core concept of the search-driven constitution-based framework is to transform evaluation requirements into generic principles that are outlined in a constitution, so these principles could be reused over time and potentially be shared across \xjudgesystems for similar \xapps, reducing manual effort and ensuring judgment quality. Developers could further customize these principles to fit the specific needs of their \xapps within their own specific context. Additionally, for a given principle, our framework searches for the most appropriate required components (e.g., cognitive architectures, jury FMs and their interactions) to construct a judging \xapp. The constitution generates qualified data points to evaluate the judging \xapp, making it an iterative process that optimizes the \xjudgesystems. To streamline the labor-intensive and ad-hoc process of evaluation, our framework divides the process into four stages: (I) creation of a general constitution, (II) specialization from general to the contextualized constitution, (III) searching for cognitive architectures using the contextualized constitution, and (IV) evolving the judge.
% (1) general principle construction, (2) specified principle construction, (3) search-driven \xapp evaluation using the specified principles, and (4) general principles refinement. 
Each stage is conducted by at least one dedicated FM-powered AI agent.  
These FM-powered AI agents are reusable components across \xjudgesystems to adapt to the dynamic changes of both \xapp and their FMs, and align with predefined \xapp requirements.   
% Furthermore, multi-agent orchestration ensures scalability and flexibility of the \xjudgesystems, especially with large datasets and large-scale \xapp, with each agent overseeing the others to ensure the quality of the judgments. 

Our paper makes the following contributions: 
\begin{enumerate}
    % \item systematically introduce the development life cycle of \xjudgesystems;
    \item Discuss nine challenges associated with developing \xjudgesystems and highlight their implications on \xjudgesystems;
    \item Propose a search-driven constitution-based framework to improve the productivity of developing \xjudgesystems and their quality;
    \item Showcase that the majority of general principles can be reusable across \xjudgesystems and the accuracy of the \xjudgesystems developed with our framework outperforming the \xjudgesystems developed without our framework by up to 6.2\%.
\end{enumerate}

\section{Challenges of Developing \xjudgesystems using a Motivational Example}
\label{sec:lifecycle}
Although existing related work touches on the issues of developing \xjudgesystems and suggests ways to improve them, there is a lack of systematic thinking about the engineering of \xjudges throughout their life cycle. Since \xjudgesystems are a special type of \xapps and face the typical challenges in the life cycle of \xapps~\cite{hassan2024rethinking}, we highlight the challenges that hinder the development of \xjudgesystems and affect the quality of \xjudgesystems significantly in this section. 
% In this section, we introduce the life cycle of the construction of \xjudgesystems and discuss the challenges developers face in each of the stages in the life cycle. This comes from our industry experience and is not meant to be comprehensive but serves as a starting point for the conversation. Since \xjudgesystems are a special type of \xapps and face typical challenges in \xapps~\cite{hassan2024rethinking}, we highlight the challenges that hinder the development of \xjudgesystems and affect the quality of \xjudgesystems significantly in this section. Fig.~\ref{fig:lifecylce} presents a typical life cycle of AI judging engineering for evaluating \xapps.
%and Table 1 presents the challenges grouped by the stages. 
We introduce a motivational example in Section \ref{subsec:mot-example} and describe the challenges along with the example. The challenges are categorized into three groups: 1) Defining judging requirements in Section \ref{subsec:challenges-defining-requirements}, 2) Developing judges in Section \ref{subsec:challenges-developing}, and 3) Evolving judges in Section \ref{subsec:challenges-evolving}.

\subsection{A motivational case}
\label{subsec:mot-example}
We select the commit message generation (CMG) task~\cite{wu2024commit, shi2022race, mcmd_dataset, dong2022fira} as a motivational case to elaborate the challenges around the development of \xjudgesystems. The CMG task involves writing concise and descriptive commit messages for a given code diff in natural language. The commit messages are important for program comprehension and maintenance and collaboration among developers.

Below we elaborate on the challenges of developing a judge system, in the context of judging a CMG system.

\subsection{Challenges associated with defining judging requirements}

\label{subsec:challenges-defining-requirements}

\noindent \textbf{B.1 Complexity of articulating requirements in a concise and clear manner.}
% \noindent \textit{Description.} 
% The stage of determining judging requirements is complex, difficult, and ad-hoc for developers. 
A primary complexity arises from the probabilistic nature of the outputs that are generated by \xapp and its embedded FMs, rather than the deterministic outputs in conventional software applications. This difference complicates the task of defining precise requirements. Developers usually define ambiguous requirements specific to their use cases, such as faithfulness to contextual information, or the naturalness of conversation. To address the ambiguity, developers then refine these requirements to various detailed dimensions and criteria. This iterative process allows developers to better align their goals with evaluating the capabilities of the \xapp and typically involves testing the performance of the \xjudgesystem to ensure that the outputs meet their expectations.

In the context of CMG, developers will need to iteratively modify requirements until they arrive at a clear and concise version. Developers may initially come up with an ambiguous requirement such as ``\textit{the commit message should be clear and concise to the changes made in a commit}'', and expand it with more details such as \textit{``Use clear and descriptive language to convey the purpose of the commit. Avoid jargon and ambiguous terms to ensure that anyone reading the message can understand the changes made.''}, before arriving at a concise version below: 

\begingroup
\smallskip
\footnotesize
\noindent \texttt{Be Descriptive and Specific:}
\begin{enumerate}[leftmargin=0.4cm, topsep=-0.1cm, label={-}]
    \item \texttt{Clearly describe what the commit does, focusing on its impact within the codebase.}
    \item \texttt{Avoid vague messages like `fixed bugs' or `updated code'.}
    \item \texttt{Avoid jargon that might not be universally understood.}
    \item \texttt{...(Truncated)}
\end{enumerate}
\endgroup
\vspace{0.2cm}

\noindent \textbf{B.2 Low efficiency in the requirements process of an AI judge as there is limited reuse across \xjudgesystems and using general benchmarks does not work for specific contexts.}
Developers need to align the judging requirements with diverse business use cases and intended goals, as this requires a profound understanding of both the technical details of the \xapp and the specific needs of the specific context at hand. Hence, developers have to extensively collaborate with the stakeholders to capture both technical and business aspects. Developers face many challenges in the communication to specify the requirements, such as various standardized practices and inadequate artifacts or information \cite{oran2021framework}. 

% In addition, developers need to meet intended goals and use cases of the target \xapp from both business and technical aspects. 

In the context of CMG, developers need to specify requirements for different programming languages since they have different standards for commit messages. For example, the C++ standard changes should be recorded in the commit messages. The use or modification of streams and lambda expressions should be highlighted as these features are pivotal in Java 8+. This is a time and labor consuming process for developers and worse the ability to reuse parts of requirements across different \xapp contexts is very low.
% \jt{double check the comment}.

%  For CMG, developers cannot use any general benchmarks (e.g., MMLU~\cite{hendrycks2020measuring}, GSM8K~\cite{cobbe2021training}) since they are not designed for the use case. Although developers gain experience of developing an \xjudgesystem for CMG, the requirements for the CMG are barely reused for another task, such as writing code comments.  

\subsection{Challenges associated with developing \xjudgesystems}
\label{subsec:challenges-developing}

\noindent \textbf{C.1 The difficulty of judging fosters continuous and rapid innovation with many new ideas coming up for architectures, metrics, and judge-optimized FMs.} Developers need an approach to select appropriate judging architectures for the predefined criteria and methods before the implementation, since
% according to the predefined criteria and metrics. 
judging involves complex steps, including selecting jury FMs, metrics and architectures, deriving judging cognitive architectures, and integrating them seamlessly. Deciding the most appropriate jury FMs in \xjudgesystems poses challenges due to the high cost and diverse judging requirements. Intuitively, developers usually select a strong FM (e.g., GPT-4) as the jury FM, since a strong jury FM obtains more accurate and reliable judgments. However, using strong models is usually expensive~\cite{gpt4expensive2023hackernews, gpt4expensive2024openai} and they are biased to the outputs generated by their model families~\cite{gallegos2024bias}. Alternatively, researchers and practitioners obtain open LLM leaderboards (e.g., HuggingFace open LLM leaderboard~\cite{hfllmleaderboard}) and arenas (e.g., Chatbot arena~\cite{chiang2024chatbot}, Alpacafarm~\cite{dubois2024alpacafarm}) where developers could see the performance and rankings of jury FMs and use it for the selection. However, the rankings of models obtained on the open leaderboards are derived from common benchmarks that are less likely to meet the specific business requirements of target \xapp for a particular context. Developers usually realize such limitations after at least a round of evaluation.

\begin{figure}[t]
  \centerline{\includegraphics[width=\linewidth]{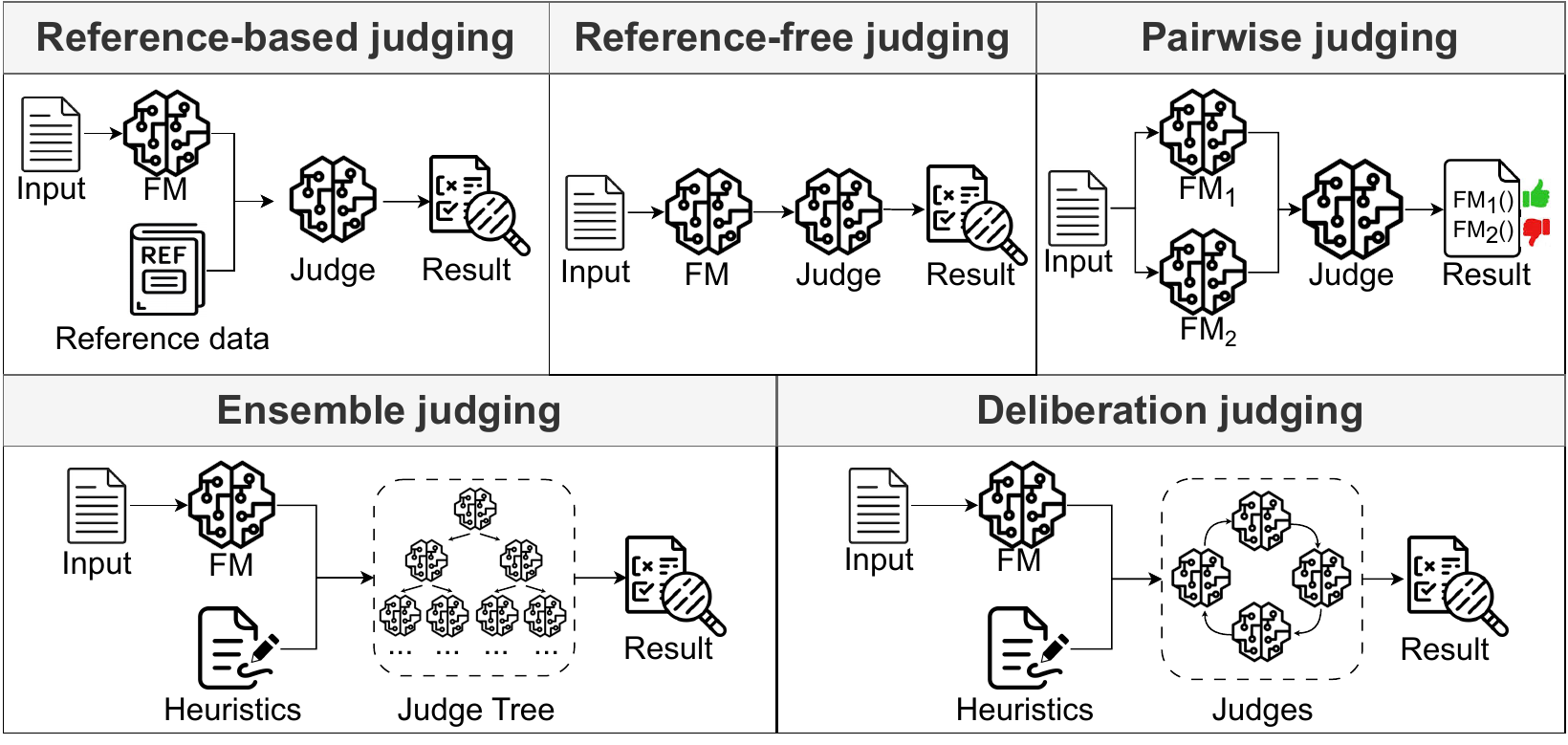}}
  \caption{Illustration of common cognitive architectures for judging.}
  \label{fig:evaluation}
\end{figure}

Developers either leverage common judging cognitive architectures or derive their own evaluation methods in their \xjudgesystems. Common cognitive architectures for judging~\cite{khan2024debating,  kenton2024scalable,  thakur2024judging, msft-agent-metrics-llm-judges, irving2018ai} include reference-based judging, reference-free judging, pairwise judging, ensemble judging, and deliberation judging, as shown in Fig \ref{fig:evaluation}.
% and human-in-the-loop (Fig \ref{fig:evaluation}).\dayi{human-in-the-loop not mentioned in either the diagram or the text. Drop?} 
\textit{Reference-based judging} requires the ground truth that is derived by deterministic functions or real-world data. This judging architecture assesses how well the generated output aligns with the ideal response. 
% For example, BLUE, ROUGE, and METEOR are common evaluation methods used to evaluate the quality of FM-generated text, but they mainly focus on the lexical level features, such as word overlap, instead of capturing the semantic meaning of a text.
% \dayi{not related to AI judge.. any non-static-metric based method for examples here? Or just give a sample of judging prompt for reference based judging} 
By contrast, \textit{reference-free judging} evaluates the responses of an \xapp without any ground truth, and relies on the internal knowledge of the jury FM and learned patterns within the prompt to determine the quality. 
% This judging architecture is more flexible and can handle a wide range of responses better but might lack the consistency of reference-based judging.\dayi{do you have citation for this? Otherwise just describe the method and don't say which one is better} 
\textit{Pairwise judging} allows a jury FM a more granular and direct evaluation of two instances against each other, leading to more reliable and consistent judgments, compared to single instance evaluation~\cite{thakur2024judging, chen2024humans}. This judging architecture is widely used in scenarios where subjective judgment is required, such as the relevance of retrieved documents. 

The \textit{ensemble judging} involves several \xjudges that issue their own decision and reach a conclusion based on certain rules (e.g., hierarchical structures, majority votes~\cite{verga2024replacing}). Such a judging architecture can decompose evaluation criteria into binary or categorical decisions, which is useful for NLP tasks like content moderation. \textit{Deliberation judging} involves multiple jury FMs engaging in structured debates to evaluate complex outputs, such as text quality~\cite{chan2023chateval} or translation~\cite{liang2023encouraging}. This architecture captures diverse perspectives and mitigates biases, leading to more balanced and thorough evaluations. Kenton et al.~\cite{kenton2024scalable} reported that debate generally outperforms consultancy and direct question-answering, especially in tasks with information asymmetry, and ensures weaker judge accuracy. While the ensemble judging architecture is efficient and transparent, the deliberation architecture offers depth and bias reduction, making both architectures complementary in ensuring the robustness and fairness of \xjudgesystems.

% \smallskip \noindent \textit{Challenges.} 
% Deciding appropriate judging models in \xjudgesystems poses challenges due to the high cost and diverse evaluation requirements. Using strong models is usually expensive~\cite{gpt4expensive2023hackernews, gpt4expensive2024openai} and they are also biased to the output generated by their model families~\cite{gallegos2024bias}. When developers want to select a proper judging model from the open leaderboards, similar to using common benchmarks, the rankings of models obtained on the open leaderboards are derived from common benchmarks that are less likely to meet the specific business requirements of the target \xapp. Developers usually realize such limitations after at least a round of evaluation and are forced to obtain private leaderboards by themselves.
% , which is costly to perform the evaluations and maintain the leaderboards.

Developers need to implement various mitigation strategies to improve the fairness and accuracy of \xjudgesystems, since jury FMs are known to be biased and unfair due to the training data~\cite{gallegos2024bias, bender2021dangers}, even for the strong models. Prior studies have reported many biases~\cite{zheng2023judging, saito2023verbosity}, including verbosity bias, positional bias, and familiarity bias, that lead to invalid judgments. For example, the verbosity bias refers to that \xjudges preferring longer and verbose text than short ones, even if the longer ones are not as clear, high-quality and accurate as the alternatives. To mitigate the effect of the bias, developers need to implement various techniques (e.g., Chain-of-Thought (CoT), in-context learning (ICL))~\cite{saito2023verbosity} or judging architectures (Fig. \ref{fig:evaluation}). Another critical bias is positional bias, which refers to that \xjudges prefer a certain position over the others in pairwise judging. Similar to the verbosity bias, developers need to employ mitigation (e.g., swap the comparisons) for such a bias. In other words, developers require considerable effort to aggregate a wide range of mitigations into their judging architectures to mitigate all the biases. Additionally, there might exist new biases that have not yet been discovered, so \xjudges exhibit unreliable and inaccurate results and developers need to fix them on a case-by-case basis.

Another key challenge is inconsistent judgments even when using identical prompts to one particular jury FM~\cite{wei2024systematic}. Developers may leverage well-structured prompts to guide jury FMs towards uniform decision-making in a logical manner~\cite{chen2023unleashing} or using fine-tuning techniques to reduce variability in outputs~\cite{parthasarathy2024ultimate}. 
% Developers are not able to determine the evaluation results (i.e., pass or fail) based on the inconsistency judgments and probabilistic behaviors, compared to the deterministic nature of conventional software applications. 
Inconsistent judgments originate from the probabilistic nature of FMs~\cite{chen2022towards} and can undermine the reliability and fairness of evaluations, making it difficult to trust the judgments. Again, developers must employ mitigation and the mitigation may differ across different kinds of judging tasks.

Developers need to create diverse judging metrics for various judging requirements, since automatic metrics primarily focus on surface-level similarities~\cite{wu2024commit} and may overlook the contextual relevance, coherence, and creativity of the generated text. 
Consequently, researchers and practitioners leverage the instruction understanding and reasoning abilities of FMs to derive diverse methods by prompt engineering, i.e., FM-based metrics~\cite{gao2024llm}. In this case, developers spend considerable time crafting prompt templates for every FM-based metric as one is unlikely to arrive at a perfect set of prompt templates in a single attempt, i.e., suffer from challenges when tuning prompts~\cite{khattab2023dspy}. Furthermore, jury FMs might not adhere to the instructions due to the inherent limitations. For instance, jury FMs are known to struggle with numerical data~\cite{chang2024survey} and are not able to assign negative scores~\cite{desmond2024evalullm}. 
% The survey in the work \cite{desmond2024evalullm} reported the need of defining negative scores to identify extreme cases in evaluating the quality of generated text.

% \noindent \textit{Example.} 
In the context of CMG, developers need to know how to make accurate judgments for the requirements. For example, given the requirement mentioned above \textit{``Be Descriptive and Specific''}, developers are not able to use any automatic metrics and decide to use FM-based metrics. In such a context, developers initially need to decide which model should be used as the jury FM, such as choosing Claude 3.5 Sonnet vs. GPT-4o. They may look at the performance of FMs from open LLM leaderboards but there are not any benchmarks regarding CMG. Developers may select Claude 3.5 Sonnect since it outperforms GPT-4o in code generation tasks\footnote{https://blog.getbind.co/2024/06/21/claude-3-5-sonnet-does-it-outperform-gpt-4o/}. In terms of metrics, developers may write a prompt to the jury FM to assign a score for a given commit message. These metrics could include 1) a raw score: representing a numerical evaluation based on predefined criteria; 2) a ranging metric: categorizing performance within specific ranges (e.g., excellent, good, average, and poor); or 3) a binary metric: providing a simple pass/fail outcome~\cite{gao2024llm}. An example of a prompt template is presented: 

\begingroup
\smallskip
\footnotesize
\begin{enumerate}[leftmargin=0cm, topsep=-0.1cm, label={}]
    \item \texttt{Given a code diff and a commit message. Assign a score range from 0 and 1 to the requirement:}
    \item \texttt{Code diff: \{\{code-diff\}\}}
    \item \texttt{Commit message: \{\{commit-message\}\}}
    \item \texttt{Requirement: \{\{requirement\}\}}
    \item \texttt{Score: }
\end{enumerate}
\endgroup
\vspace{0.2cm}

% \noindent \textbf{2) The field evolves rapidly, so developers have to always learn new advancements to integrate them.}
% Developers are in a constant race to keep up, integrating the latest innovations to maintain cutting-edge performance when they are developing their \xjudgesystems. This relentless pace of development demands flexibility and adaptability from developers to stay ahead in the competitive landscape.

% In the context of CMG, developers may observe a new FM is released and need to explore whether the new FM performs better than the used one. For example, OpenAI o1-mini was released on Sept 12th, 2024\footnote{https://openai.com/index/openai-o1-mini-advancing-cost-efficient-reasoning/} and the model excels in certain areas (e.g., coding). Therefore, developers need to analyze the cost and performance of o1-mini vs. Claude 3.5 sonnet and decide whether they need to switch to o1-mini. Once they decide to use o1-mini, they have to integrate ol-mini into the current \xjudgesystem and review the performance. 

\medskip \noindent \textbf{C.2 No silver bullet method instead mostly case-by-case optimization.} There is no universal solution or best practices that fit all judging tasks. Optimization strategies need to be tailored to the specific requirements of each use case. This means developers must carefully analyze the unique aspects of each judging task and apply appropriate components in their judging architecture. Given the diverse judging architectures with each of them having limitations, the complexity of setting up and maintaining the judging architectures can be high, and there might exist nuanced or ambiguous cases that do not fit neatly into predefined categories. For example, the pairwise judging might be less feasible for online evaluation since it is less likely to find a proper comparison instance for every user query. Developers might struggle with balancing the trade-offs between accuracy and resource limitations~\cite{wilkins2024offline}. 
% They may find it challenging to select the most effective heuristic or judging architecture for a particular task, especially when faced with varying data quality and context. 
Designing judging architectures typically leads to an ad-hoc version of the \xjudgesystem that is designed for one particular \xapp, i.e., labor burden.

In the context of CMG, developers do not know how to optimize the \xjudgesystem since FMs have their weaknesses and different judging architectures fulfill their specific contexts. For example, using the reference-free judging architecture vs. the decision-tree judging architecture. While the reference-free architecture is simple to develop, the decision-tree judging architecture is more complex, but a more nuanced and robust decision-making process. Developers need to develop both architectures and verify which one is better.

\medskip \noindent \textbf{C.3 Creation of judging data is labor intensive.} Developers typically select datasets by three common strategies: (1) leveraging common benchmarks (e.g., MMLU~\cite{hendrycks2020measuring} and GSM8K~\cite{cobbe2021training}), (2) manually curating a dataset, and (3) synthesizing a particular dataset. The first strategy is often employed when evaluating FMs, while the others could be used in the evaluation of both FMs and \xapps. However, none of the aforementioned strategies is perfect. Common benchmarks are designed to evaluate the general capabilities (e.g., math) of FMs and may not be specific to a given set of requirements. In addition, the benchmarks may have leaked to FMs during pretraining~\cite{balloccu2024leak}, ending up with invalid evaluation results. Manually curating datasets is both time-consuming and labor-intensive~\cite{liu2024best}. Despite the availability of various synthetic data generation techniques (e.g., ZeroGen~\cite{gao2022self}, SuperGen~\cite{meng2022generating}, Alpaca~\cite{taori2023alpaca}), synthetic datasets present challenges in areas such as factuality and fidelity of synthetic data~\cite{wood2021fake} and often fail to generalize to real-world scenarios~\cite{van2023synthetic}. Additionally, after a dataset has been curated, developers need to maintain the dataset to be applicable when FMs and the \xapp co-evolve.

In the context of CMG, developers need to specify test datasets for different levels and types of changes across different programming languages to ensure the test coverage of the \xjudgesystem. Developers could make a code diff at the line, class, method, package, or system level and the diff could be associated with any of the following types: bug fixes, refactoring, feature additions, performance optimization, and security updates. The creation of test data for all the aforementioned changes for all programming languages is labor-intensive.

\medskip \noindent \textbf{C.4 Balancing between cost, accuracy, and latency is crucial.} Developers often prioritize the accuracy of \xjudgesystems over the other factors due to its critical importance. However, to achieve high accuracy, developers often leverage techniques such as calibrating prompts with examples to a jury FM to reduce its weakness~\cite{zheng2023judging}, i.e., increase costs due to the increased number of input tokens, or increasing the number of requests to jury FMs~\cite{verga2024replacing}, i.e., increase latency. Balancing these factors requires a nuanced understanding of judging cognitive architecture and a flexible approach to adapting new techniques.

In the context of CMG, developers may aim to improve the accuracy of the \xjudgesystem using few-shot learning. To achieve this, developers incorporate examples of code diffs and their corresponding commit messages into the prompt, providing substantial context in the prompt to guide the jury FM. While this approach improves the accuracy of judgments, it significantly increases the cost due to the high token count.   

\subsection{Challenges associated with evolving \xjudgesystems}
\label{subsec:challenges-evolving}

\noindent \textbf{D.1 In the rapidly active field of \xapp, developers always have to chase new techniques and update them to the evolving \xjudgesystems.} The relentless pace of innovation in the field means developers must constantly adapt to incorporate the latest advancements. The continuous need for integration can be overwhelming, requiring frequent updates and adjustments to ensure the \xjudgesystem remains at the cutting edge. In addition, novel advancements could be raised from any unit inside an \xjudgesystem, including code, sophisticated judging architectures, FMs, and mitigation, since the change of one unit affects the overall reliability and robustness of the \xjudgesystem. For example, when a new mitigation for the positional bias is introduced, developers observe that the new mitigation outperforms the prior mitigation when testing the \xjudgesystem. However, after deploying it online, developers may discover that the new mitigation performs worse than the old one on certain real-world data subsets. Consequently, developers maintain both mitigations for different data groups. Tracking the versions of each mitigation and their usefulness against a certain bias is critical in maintenance and evolution. In addition, mitigation might not be applicable for all FMs, indicating the complexity and difficulty in the evolution of \xjudgesystems.

% For example, several approaches focused on optimizing FM inference speed, such as tailoring hardware architectures for FMs ~\cite{huang2024new}, the workload of FM by capturing the relationship between the number of input and output tokens~\cite{wilkins2024offline}, were released. 

In the context of CMG, imagine a new judging cognitive architecture is released that significantly improves the accuracy of weaker jury FMs. Developers must rapidly understand the new architecture, implement, and test the new architecture within the existing \xjudgesystem. Developers need to carefully assess the performance of the new architecture by applying all historical online data before deploying a new version to the production environment. 
% \dayi{sounds very similar to C2}

\medskip \noindent \textbf{D.2 Field problems.}
Deploying the \xjudgesystems in production environments presents numerous practical challenges. Developers must tackle a range of issues, from handling diverse and unpredictable data inputs to ensuring the robustness and reliability of the \xjudgesystem, since common pitfalls (e.g., biases~\cite{gallegos2024bias}, hallucinations~\cite{huang2023survey}, human preference alignment~\cite{wei2024systematic}) are amplified when using the real-world user data and potentially end up with negative consequences (e.g., reputational damage or incidents). One of the important challenges is data drift due to real-world data and jury FMs themselves~\cite{kirk2023understanding, chen2023chatgpt}. Real-world data often exhibits out-of-distribution (OOD) instances that were not apparent during testing. 
% Similar issues exist in practice when using AI in software engineering, such as incidents in cloud service operations due to inadequate data~\cite{kapel2024difficulty, batta2021system}. 
Additionally, jury FMs may drift significantly over a short period, especially when using proprietary models (e.g., OpenAI). Chen et al.~\cite{chen2023chatgpt} reported significant behavior changes between the March 2023 and June 2023 versions of GPT-4, including a decreased ability to follow user instructions. Similarly, developers experienced significant performance degradation while using Claude 3 Sonnet in March, 2024, a few weeks after it was released~\cite{claude32024behaviorchange}. 
These subtle changes compromise the reliability and accuracy of \xjudgesystems significantly. 
% As a result, developers address such challenges by incorporating new changes on a case-by-case basis.
% \dayi{You can specify a certain model version from model providers such as OpenAI to avoid model changes right? Also this is not limited to AI judges but any FMware relying on proprietary models?}

% Online evaluation occurs when an \xjudgesystem being deployed to production, either independently or being integrated into a \xapp, as mentioned in Section~\ref{sec:background}. Common pitfalls (e.g., biases~\cite{gallegos2024bias}, hallucinations~\cite{huang2023survey}, human preference alignment~\cite{wei2024systematic}) are amplified when using the real-world user data and potentially end up with negative consequences (e.g., reputational damage or incidents). 

% Prompt injection is another important challenge to the reliability and accuracy of online evaluation. Attackers could leverage prompt techniques (e.g., jailbreaks, adding a fake reference) to deceive the judges~\cite{perez2022ignore, chen2024humans}. Although judging models continue to be improved in defense against more prompt injections, new prompt injection techniques continue to emerge. Therefore, developers need a general approach to address such challenges and upgrade their judge systems rapidly. Privacy concerns (e.g., the exposure of sensitive data) and reproducibility are also key challenges in online evaluation. Continuous monitoring and iterative improvements are essential to maintain the integrity and effectiveness of AI judge systems in dynamic environments.

In the context of CMG, the deployed \xjudgesystems may require additional handling for certain real-world data. For example, a commit message includes sensitive information (e.g., API token) that the \xjudgesystem is less likely to identify such information and determines the low quality of the commit message.

\medskip \noindent \textbf{D.3 New/better understanding of requirements.} As the \xjudgesystem is deployed in production, developers gain new insights into its requirements. This evolving understanding can lead to significant changes in tuning jury FMs, judging architectures, feature development, and data collection strategies. While this deepening knowledge improves the system's effectiveness, it also requires developers to constantly reassess and update their judge systems \cite{pan2024human}.

In the context of CMG, developers may get new requirements that the deployed \xjudgesystem needs to support Kotlin, a preferred programming language on Andriod development. Developers need to explore issues such as how to emulate an Android environment within the \xjudgesystem and integrate the environment seamlessly. 
\section{Search-driven constitution-based framework for AI judge systems}
\label{sec:framework}

\begin{figure}[t]
  \centerline{\includegraphics[width=\linewidth]{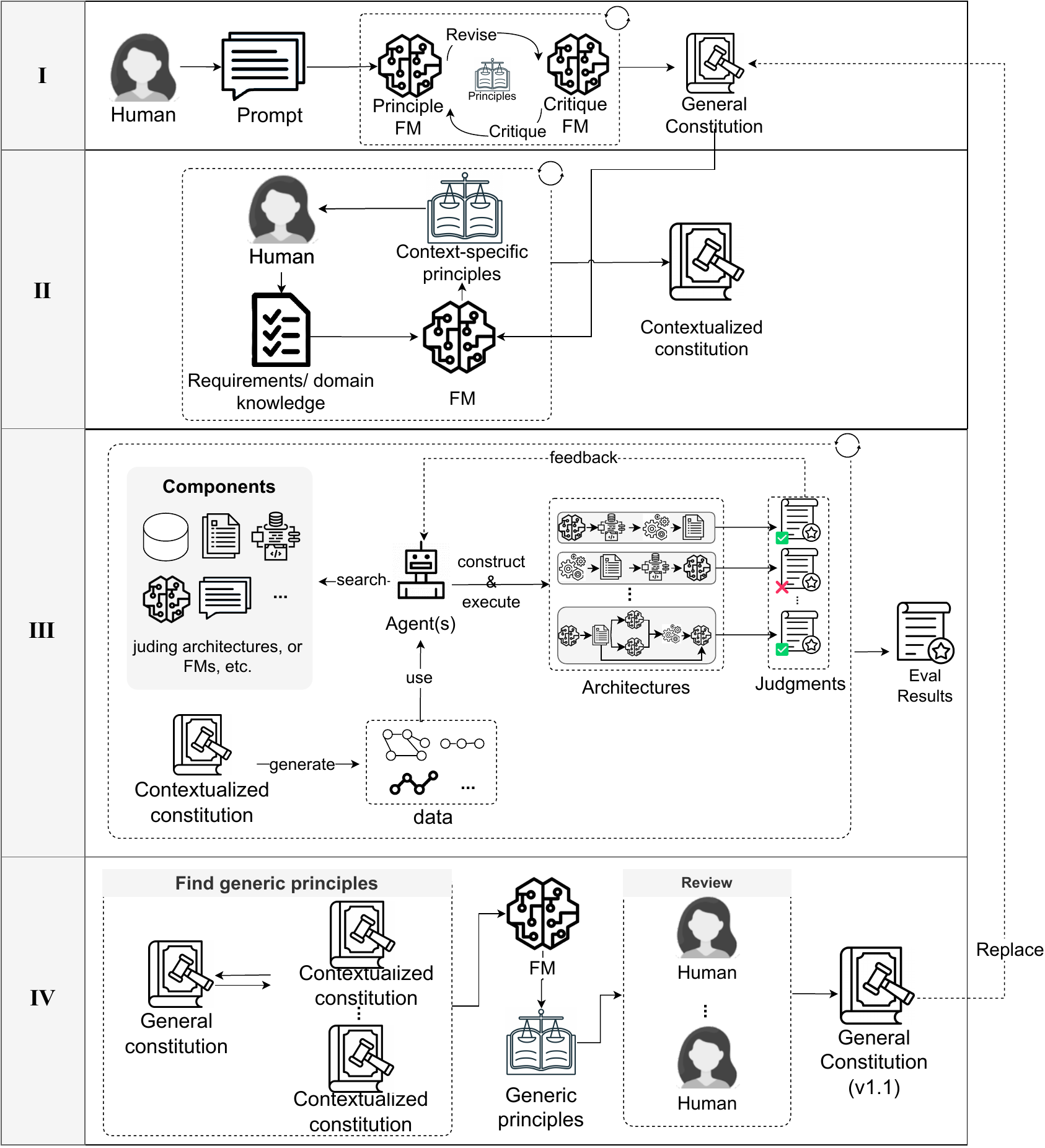}}
  \caption{Overview of the search-driven constitution-based framework. The framework includes a total of 4 stages: I) creation of a general constitution, (II) specialization from general to the contextualized constitution, (III) searching for cognitive architectures using the contextualized constitution, and (IV) evolving the judge.}
  \label{fig:framework}
\end{figure}

In light of the discussed challenges in the prior section, we aim to increase the productivity of developing \xjudgesystem while ensuring the quality of judgments through 
% continuously learn the dynamic changes of both an \xapp and its FMs overtime and to be capable of continuously evaluating both of the FMs and \xapp 
a search-driven constitution-based framework (Fig.~\ref{fig:framework}). In an \xjudgesystem, the principles outlined in a constitution are designed to specify and quantify judging requirements, along with their associated criteria and methods. The concept of constitutions was introduced in a prior work~\cite{bai2022constitutional} where the authors designed a set of guiding principles in a constitution that models must adhere to ensure they are harmless, helpful, and honest.
The framework transforms requirements into principles as knowledge in \xjudgesystems, i.e., from a \textit{data-driven} approach to a \textit{knowledge-driven} approach. Since the \textit{data-driven} approach often exhibits inefficiencies in data utilization and lacks precise control mechanisms~\cite{aldoseri2023re}, while the \textit{knowledge-driven} approach uses a graph-structured data model or topology to represent and operate on knowledge~\cite{kejriwal2019domain, lecue2020role}. 
% For example, Google announced that using knowledge graphs improved the results of Google Search significantly~\cite{intro-knowledge-graph}. 
% \dayi{briefly define relevant concepts such as constitutions, principles, knowledge etc. here, and motivate why constitution driven approach is better than data driven approach} 

The framework includes the following stages: (I) creation of a general constitution, (II) specialization from general to the contextualized constitution, (III) searching for cognitive architectures using the contextualized constitution, and (IV) evolving the judge. 
% The framework employs multi-agent orchestration that allows agents to supervise one another, thereby ensuring the overall quality of the \xjudgesystem. 
We detail the stages in the following subsections. 

% \dayi{I think we should point out which specific challenge is each of the following stage addressing, in each stage's description.}

\subsection*{Stage I. Creation of general constitution}
\label{subsec:framework-stage1}

The primary goal in this stage is 
% The primary objective of generating a set of principles is 
to transform the requirements into general and reusable guidelines that remain applicable, even when the \xapp evolves and when data and models drift over time, aiming to address challenges B.1, and B.2. The framework transforms judging requirements into criteria in a certain context. Such transformation in
% among requirements, criteria, and metrics as principles
the framework provides a consistent and scalable method for the ongoing evaluation and potential changes in the \xapp that developers of the \xjudgesystem could leverage when maintaining and updating the \xjudgesystem. 

% to ensure reliability, robustness and flexibility of the \xjudgesystem. 

% The goal is to transition from a data-driven approach to a knowledge-driven (knowledge graphs) approach, leveraging the principles in the public constitution to facilitate the evaluation process, since data-driven AI often exhibits inefficiencies in data utilization and lacks precise control mechanisms~\cite{aldoseri2023re}, while a knowledge graph uses a graph-structured data model or topology to represent and operate on data~\cite{kejriwal2019domain, lecue2020role}. For example, Google announced that using knowledge graphs improved the results of Google Search significantly~\cite{intro-knowledge-graph}.

% This hierarchy facilitates the ongoing evaluation and potential changes in the \xapp while maintaining alignment with foundational principles. Thus, the framework supports robust and flexible evaluation processes that can adapt to dynamic changes.

% Our framework uses a top-down approach to convert judging requirements, criteria and metrics into principles. 

First, each predefined judging requirement is transformed into a set of principles (i.e., \textit{requirement-principle}) through prompts given to an FM. These principles are refined through at least four rounds of critiques and revisions, as suggested by the prior work~\cite{bai2022constitutional}, to ensure quality and applicability. Further optimization of these principles can be achieved using techniques such as CoT, few-shot learning, or fusion techniques. In the context of CMG, one example of the requirements is defined as \textit{``clarity, conciseness, and relevance to the changes made in the commit''}. The corresponding example of the \textit{requirement-principles} for clarity is presented:

\begingroup
\smallskip
\footnotesize
\begin{enumerate}[leftmargin=0cm, topsep=-0.1cm, label={}]
    \item \texttt{Use clear and descriptive language to convey the purpose of the commit. Avoid jargon and ambiguous terms to ensure that anyone reading the message can understand the changes made.}
\end{enumerate}
\endgroup
\vspace{0.2cm}

Next, for each \textit{requirement-principle}, a corresponding set of principles is generated 
% through additional prompting of FMs 
and these principles are served as evaluation criteria (\textit{criteria-principles}). This process also involves multiple rounds of critiques and revisions to refine the criteria-principles, resulting in a final set of criteria-principles. 

% Similarly, metrics are derived from the criteria-principles through a comparable iterative process, leading to the establishment of \textit{metric-principles}. 
% One example is presented as follows:

These sets of principles collectively form a general constitution, that is embedded into an agent's knowledge. Such an agent could serve as 
% \dayi{rephrase here, because constitution cannot be an agent, they are different concepts. maybe "... public constitution, that is embedded into an agent's knowledge. Such an agent can serve as..."} that serves 
two roles: validator and advisor. The validator role monitors and assesses changes in the evolution of an \xapp and its FMs, ensuring that updates adhere to the established principles. The advisor role provides guidance to developers of \xjudgesystems, helping them align their evaluations with the core principles and maintain consistency across different versions of the \xapp.

The general constitution is designed to be reused across various developer teams in the next stage. These teams work on building an \xjudgesystem for similar types of \xapp, such as text summarization, but across different contexts or dimensions, such as technical documents vs. news. This adaptability allows the general constitution to serve as a flexible and reusable resource, supporting diverse evaluation needs in the same foundation and fostering consistency across different \xapp.

\subsection*{Stage II. Specialization from general to the contextualized constitution}
\label{subsec:framework-stage2}
% [knowledge graph?]
% \jt{Take Constitution to generate the data to run the search}
The primary objective of this stage is to incorporate context-specific knowledge into the constitution, develop a new set of specific principles with assistance from an FM that embeds the general constitution as knowledge, and form a new constitution. This stage is designed to address the challenges B.1, B.2, and C.3. This process involves collaboration between developers and the constitution to establish context-specific sets of principles.
% , including \textit{req-principles}, \textit{criteria-principles}, and \textit{metric-principles}. 
% The goal is to transition from a data-driven approach to a knowledge-driven (knowledge graphs) approach, leveraging the principles in the public constitution to facilitate the evaluation process, since data-driven AI often exhibits inefficiencies in data utilization and lacks precise control mechanisms~\cite{aldoseri2023re}, while a knowledge graph uses a graph-structured data model or topology to represent and operate on data~\cite{kejriwal2019domain, lecue2020role}. For example, Google announced that using knowledge graphs improved the results of Google Search significantly~\cite{intro-knowledge-graph}.

In this stage, developers prompt the FM using any prompt engineering techniques (i.e., RAG, ICL) to get a new set of context-specific principles tailored to their subjective requirements. The developers then review and critique these new principles (either manually or by an FM), and request refinements from the FM that embeds the general constitution as knowledge. Similar to Stage I, this interactive process involves several rounds of feedback and adjustment until a final and well-satisfied set of principles is obtained. The final set of context-specific principles forms a contextualized constitution that incorporates the context-specific knowledge and expertise relevant to the target \xapp. In the context of CMG, one defines context-specific principles for writing commit messages for a given diff written in C++, along with its detailed criteria as follows:

% \begingroup
% \smallskip
% \footnotesize
% \noindent \texttt{Be Descriptive and Specific:}
% \begin{enumerate}[leftmargin=0.4cm, topsep=-0.1cm, label={-}]
%     \item \texttt{Clearly describe the commit and include affected files or modules. Use precise C++ terminology such as classes, methods, functions, or libraries.}
%     \item \texttt{Avoid overly technical jargon that could be unclear to the broader team.}
%     \item \texttt{Example of a good commit message: Fix memory leak in `DataProcessor::releaseMemory()' by ensuring correct `delete' usage.}
%     \item \texttt{...(Truncated)}
% \end{enumerate}
% \endgroup
% \vspace{0.1cm}

\begingroup
\smallskip
\footnotesize
\noindent \texttt{Mention C++ Standard or ABI Changes:}
\begin{enumerate}[leftmargin=0.4cm, topsep=-0.1cm, label={-}]
    \item \texttt{If changes depend on specific C++ standards (e.g., C++17, C++20), mention the standard in the commit message.}
    \item \texttt{If ABI stability is impacted, such as when upgrading libraries or changing compilers, make this explicit.}
    \item \texttt{Example: Refactor `FileHandler' to use `std::filesystem' (requires C++17).}
\end{enumerate}
\endgroup
\vspace{0.1cm}

Moreover, developers could leverage the contextualized constitution to curate a test dataset, either by reviewing and refining an existing dataset to identify and correct inappropriate data points, or generating synthetic data from a few seed manually created data points. The curated dataset could be continuously reviewed and refined, when the \xapp evolves and when the data and models drift. As such, the curated dataset remains relevant to the target \xapp and reliable in the evaluation process, thereby ensuring the accuracy and reliability of judgments made by the \xjudgesystems.  

% Developers could continuously manage and maintain the curated dataset  

% Such an approach helps developers manage and maintain datasets effectively, especially as an \xapp evolves and as data and models drift over time. By leveraging a well-defined set of principles, developers can ensure that the test datasets remain relevant and reliable when evaluating the target FMware, thereby ensuring the accuracy and reliability of judgements made by the AI judge system. 

\subsection*{Stage III. Searching for cognitive architectures using the contextualized constitution} 
\label{subsec:framework-stage3}
% [cite prior work and talk about the eval architecture, granularity, dynamic]
% \dayi{check SE3.0 for searching.. constitution needs to be used to generate data points, which is used for the optimization loop of architecture search. And I don't get the knowledge graph part.}
% \jt{highlight the search agent}
The primary objective of this stage is to facilitate the development of \xjudgesystems while ensuring the delivery of high-quality judgments through a search-based exploration. This stage aims to address the challenges C.1-2, C.4, and D.1-2. Given a context-specific principle, Fig \ref{fig:framework} depicts that the agent searches for the optimal combination of the components (e.g., jury FMs) to construct a judging architecture. The contextualized constitution generates judging data points, which are then injected into the architecture to verify the accuracy of the judgments. This methodology draws inspiration from Test-Driven Development (TDD) in conventional software applications, which ensures that developers thoroughly understand the requirements before they start to write code~\cite{beck2002test}. The test-driven approach has been used in certain techniques, such as DSPy~\cite{khattab2023dspy}. When the agent encounters inaccurate judgments, the agent adapts by re-constructing a new architecture. This iterative feedback loop ensures that the agent continuously learns from weaker areas and refines the constructed architecture. By consistently refining the architecture, the framework enhances the overall effectiveness and precision of the \xjudgesystem. This dynamic and adaptive approach not only ensures that the \xjudgesystem remains responsive to evolving requirements but also improves the reliability and accuracy of judgments over time. Furthermore, the approach mitigates the challenge of the ever-changing field, as one can simply add new techniques to the search space of the framework, which can then automatically reconstruct a judging architecture with the updated best combination.

\subsection*{Stage IV. Evolving the judge}
\label{subsec:framework-stage4}
The purpose of this stage is to address flaws in the principles outlined in the general constitution (Stage I). This stage is designed to address the challenge B.2 and D.3. Since these general principles were derived from the judging requirements, any potential flaw should be traced back through the transformation obtained in Stage I to ultimately reveal whether there is a bug in the original judging requirement, due to the knowledge-driven approach. This is important for debugging and bug fixing in the ongoing evolution of FMs, \xapp, and \xjudgesystems. 
Our framework seeks to identify potential requirement bugs in the general constitution through a semi-auto process among the general and contextualized constitutions and the developers. Since the context-specific principles that are outlined in the contextualized constitutions have been manually reviewed and verified, the contextualized constitutions identify potential bugs or discrepancies in the general principles automatically, instead of by the developers themselves manually.
% semi-auto process assist developers identify potential bugs or discrepancies in the requirements of the public constitution and reduce manual effort.
% reduce developer effort  interaction allows developers to identify potential bugs or discrepancies in the requirements of the public constitution. 
The process involves four key steps:

\begin{enumerate}[leftmargin=0.4cm, label={•}]
    \item Identification of potential flaws: The general and contextualized constitutions perform a comparative analysis of their principles and detect potential requirement bugs by the inaccurate judgments obtained in the prior stage. This step leverages the manually verified principles from contextualized constitutions to assess and validate the principles of the general constitution. 
    \item Review and consensus: Any potential requirement bugs must be reviewed, discussed and agreed by a majority of the constitutions. This consensus-driven approach ensures that the identified requirement bugs are valid and relevant across different contexts within the \xapp.
    \item Private knowledge review: Since the contextualized constitutions often contain private knowledge or context that should remain confidential, the developers of these contextualized constitutions must review the identified potential requirement bugs.
    % developers must be further reviewed by the framers (developers) of these specific constitutions. 
    This step ensures that sensitive or proprietary information is protected while the identified requirement bugs are validated.
    \item Incorporation of fixes: After the requirement bugs have been validated and reviewed, the corrections are integrated into the general constitution. This step ensures that the general constitution evolves to address identified requirement bugs, maintaining its relevance and effectiveness.

\end{enumerate}

Our framework ensures that the general constitution remains adaptive and robust when facing the dynamic changes in \xapps and evolving requirements.

\section{Case study and results}
\label{sec:case_study}
As discussed in Section \ref{sec:lifecycle}, we select the commit message generation (CMG) task~\cite{wu2024commit, shi2022race, mcmd_dataset, dong2022fira} as a case study to evaluate the productivity and quality of judgments made by our proposed framework. We focus on the quality of the \xjudgesystem when using reusable constitutions and their principles (Stages I and II). The aspects of searching for the best cognitive architectures and improving the general principles (Stages III and IV) will be addressed in future work. 
% \dayi{we should highlight we only work do the constitution / requirement refinement part of evaluation and search will be explored in future work}

% The CMG task involves writing concise and descriptive commit messages for a given code diff in natural language. The commit messages are important for program comprehension and maintenance and collaboration among developers.

% since it involves writing descriptive commit messages for a given code diff. Such a task serves as a proxy to demonstrate the quality of judgements in both NLP and software engineering (SE) aspects. Specifically, the CMG task would require the understanding of a code diff to generate coherent and meaningful commit messages.

\subsection{Dataset}
We employ the multi-language commit message dataset (MCMD)~\cite{mcmd_dataset} with 5 popular programming languages, i.e., C++, C\#, Java, Javascript, and Python, which has been widely used in the CMG-related studies~\cite{shi2022race, dong2022fira, shi2022evaluation, wu2024commit}. The MCMD dataset was collected from the top 100 most-starred repositories in GitHub and each data point in the MCMD dataset included a code diff and the corresponding commit messages. We leverage the MCMD dataset obtained by Shi et al.~\cite{shi2022race} that the authors added more fine-grained code change actions (e.g., $<$REPLACE\_OLD$>$) in a given code diff. 
% Furthermore, the authors added action tokens (e.g., $<$REPLACE\_OLD$>$) to suggest more fine-grained span-level code changes. 
% Due to resource constraints, 
Since the MCMD dataset contains a total of approximately 100K data points and making judgments for such a large number of data is resource consuming, we select a representative random sample set (confidence level = 95\%, confidence interval = 5\%) of data points in each of the 5 programming languages and report the corresponding sample sizes in Table~\ref{tab:basic-stats}. 

To emulate the outputs of a target \xapp, we replicate a prior work~\cite{wu2024commit} using an \xapp for the CMG task. To the best of our knowledge, this work was the latest work and had been published in one of the top SE venues. Wu et al.~\cite{wu2024commit} used the few-shot learning technique to generate commit messages and reported the performance of their approach increased when the number of shots increased, then dropped, due to the limited length of the context window of the used model (i.e., GPT-3.5-turbo-1106). Therefore, to avoid the limitation issue, we replicate the 16-shot result for making judgments and evaluating the performance of our framework. 
% As Wu et al.~\cite{wu2024commit} used few-shot learning to generate commit messages and reported that they needed to truncated some context of examples , 
% We show the statistics of Shi et al.'s dataset~\cite{shi2022race} and our data in Table~\ref{tab:basic-stats}.
% apply random sampling to select a subset of the testing data . The sample size was determined using a 95\% confidence level.

\begin{table}[t]
    \centering
    \renewcommand{\arraystretch}{1.1}
    \caption{The descriptive statistics of our study.}
    \label{tab:basic-stats}
    \begin{tabular}{|p{1.3cm}|r|r|r|r|} 
    \hline
    \textbf{Language} & \textbf{MCMD} & \textbf{Studied set$^1$} & \textbf{\# principles} & \textbf{Inc. acc (\%)$^2$} \\ \hline
    C++ & 20,141 & 377 & 14 & 6.1\%\\ 
    C\# & 18,702 & 377 & 15 & 6.2\%\\ 
    Java & 20,159 & 377 & 12 & 1.0\%\\ 
    Python & 25,837& 379 & 14 & 4.0\%\\
    Javascript & 24,773 & 379 & 14 & 4.5\%\\ 
    \hline
\end{tabular}
\begin{tablenotes}
    \item[*]$^1$ A representative random set with CL@95\% and CI@5\%, according to the number in its left cell.
    \item[*]$^2$ The percentages indicate the increased percentage of accuracy on judgments made by the \xjudgesystems developed with our framework, compared to those made by the \xjudgesystems developed without principles.
\end{tablenotes}
\end{table}

% \jt{why choose the CMG task and the MCMD dataset}
% \jt{intro the MCMD}

\subsection{Approach}
We evaluate our search-driven constitution-based framework for \xjudgesystems from two aspects: 1) the effectiveness of principles obtained in the constitutions, and 2) the quality of judgments. As discussed in Section~\ref{subsec:framework-stage1}, we prompt FMs to generate a set of general principles for writing commit messages in software development, before performing four rounds of critiques and revisions. We obtain 17 general principles in the constitution.

% To demonstrate our constitution-derived framework for \xjudgesystems, we start by generating the public constitution. Given the CMG task, we seed the FM to generate a set of general principles for writing commit message for software development. In the critique and revise loop, we utilize two stateless agents: a critique agent and a revise agent. We note that stateless and stateful agents does not have a significant difference, however stateless agents are cheaper considering the input length. For each iteration, we feed the critique agent with the revised principle. Subsequently, the revise agent takes in the critique and the revised principle. Then, we run the loop until no more critiques and modifications can be made to the principle. In our experiment, we run the loop 5 times.

In stage 2, we tailor the general principles to each of the 5 programming languages and generate context-specific principles. For each programming language, we prompt context-specific requirements (e.g., the name of the programming language) to the FM while iterating critiques and revisions. We review the revised principles and remove the inadequate ones to ensure the quality of the context-specific principles. We perform this process iteratively until the final set of context-specific principles is satisfied and verified. 
% Subsequently, we follow a similar procedure to that used in generating general principles. However, in this case, the critique agent and the revision agent base their critiques and revisions on the specific characteristics of each programming language. 
% We perform ad-hoc human-in-the-loop evaluation, where after the last principle revision, we go through them and filter the principles we want. 
% The semi-auto approach reduces the time required to generate a set of specified principles significantly, compared to the time needed manual generation from scratch, as suggested by a prior work~\cite{bai2022constitutional}. 
As a result, we obtain 5 high-quality contextualized constitutions (see Table~\ref{tab:basic-stats}). These constitutions guide the judges during the quality assessment of the FM-generated commit messages. Specifically, each context-specific principle represents a metric with detailed criteria.

To evaluate the quality of the judgments, we compare the judgments made by two \xjudgesystems: with (w/) vs. without (w/o) our proposed framework. 
% We then evaluate the commit messages using our framework. Specifically, we employed two judges: one with principles and one without. 
% The latter serves as a baseline for assessing the reliability of the \xjudge with principles.
To simplify the evaluation process, the \textit{search} agent sets the score range between 0 and 1 for every metric and selects a specific FM (i.e., GPT-4o-2024-05-13) as our jury FM for constructing the judging architecture.  
% For the principle-guided judge, we instructed it to evaluate commit messages based on the constitution, assigning a score between 0 and 1 for each principle. 
The \textit{search} agent computes the final score by summing the scores of all principles in the \xjudgesystem with our proposed framework. By contrast, the \xjudgesystem without our proposed framework is instructed with assigning a score
% commit messages 
on a scale from 0 to $n$, where $n$ is the number of context-specific principles used by the corresponding \xjudgesystem developed with our framework. This approach ensures both \xjudgesystems operate on a comparable scale and indicates the normalization of the scores. For a given dataset of a certain programming language, we obtain two scores (w/ and w/o our framework) for every data point.
% , consisting of a code diff and its FM-generated commit message.
% , effectively implementing a soft normalization of the scores.

% \jt{Based on the results of the paper~\cite{wu2024commit} \\
% Dataset: distilled version of MCMD~\cite{shi2022race} \\
% need to write what and how
% why choose 16-shot \\
% - 1. evaluate \# principles, reduce human effort 
% - 2. evaluate the perf of using principles vs. not using principles}

% To evaluate the quality of the judgements (i.e., the scores) made by the \xjudgesystems, 
We derive heuristics to determine the ground truth of judgments for several reasons. First, there is not any ground truth in the MCMD dataset for how many scores of a commit message should be assigned. Second, manually reviewing the score of every diff requires expertise in the 5 programming languages. Lastly, the manual review is time consuming and not scalable. 
We leverage relative comparisons of two scores for a given pair of data points ($a, b$) as a proxy to evaluate the accuracy of judgments. When the score of $a$ is larger than that of $b$, we determine $a$ is better than $b$. The ground-truth heuristic of relative comparisons is derived from automatic metrics and the majority-voting approach. Since the prior work~\cite{wu2024commit} utilized four metrics (i.e., BLEU~\cite{papineni-etal-2002-bleu}, ROUGE-L~\cite{lin-2004-rouge}, CiDEr~\cite{oliveira-dos-santos-etal-2021-cider}, METEOR~\cite{banerjee2005meteor}), we add one additional metric, i.e., BLEURT~\cite{sellam2020bleurt}, to reduce the probability of ties in the majority-voting approach. BLEURT is a parametric reference-based metric and captures deeper semantic relationships between the candidate and the ground truth, assessing both the fluency of the candidate and its effectiveness in conveying the meaning of the reference. 

$\mathcal{P}$ denotes the obtained set of paired data points and is composed of a total of $C(n,r) = \frac{n!}{(n-r)!r!}$ pairs, where $n$ indicates the number of data points and $r$ is 2.
A pair of data points is denoted as $p_i = (a ,b)$ at index $i$ of $\mathcal{P}$.
Based on the majority-voting approach, for a given pair $p_i$, $a$ is better than $b$ when at least the values of three used metrics of $a$ are larger than those of $b$. In the case that the values of certain metrics of $a$ are equal to those of $b$ (i.e., ties), $a$ is better than $b$ if the number of non-tied metrics where $a$ has larger values is greater than that of $b$. \texttt{LarCnt}$(p_i, a)$ denotes the number of non-equal metrics of $a$ with larger values than that of the other point $b$ in the $p_i$.
The ground-truth heuristic, denoted as \texttt{MetricVote}, 
% based on the majority vote approach 
for a given 
% pair of data points 
$p_i$ 
% at index $i$ of $\mathcal{P}$ 
is formulated as:
% \begin{equation}
% \operatorname{Vote}(x_1,x_2) = \sum_{i=1}^n [\operatorname{m}_i(x_1) > \operatorname{m}_i(x_2)]
% \end{equation} 
% where $x_1$ and $x_2$ are inputs, $m$ is a metric (e.g., BLUE), and $n$ is the number of metrics.
% Then, we define \texttt{MetricVote} to determine which sample of the pair is better as
\begin{equation}
\resizebox{0.9\hsize}{!}{
    \ensuremath{
        \operatorname{MetricVote}(p_i) = \begin{cases}
        a & \text{if } \operatorname{LarCnt}(p_i, a) 
        % (y_{i,0},y_{i,1}) 
        > \operatorname{LarCnt}(p_i, b) \\
        b & \text{if } \operatorname{LarCnt}(p_i, a) < \operatorname{LarCnt}(p_i, b) \\
        0 & \text{otherwise}
        \end{cases}
        }
    }
\end{equation}
where an output of $0$ indicates that all the values of the metrics of the two data points in the $p_i$ are equal.
% is the tied case as there had scores that equalled during the majority-voting approach. 
Similarly, the heuristic of which data point is better based on the judgments is denoted as 
% determine which sample in a pair is better with 
\texttt{JudgeVote} and formulated as:
\begin{equation}
% \resizebox{0.9\hsize}{!}{
%     \ensuremath{
        \operatorname{JudgeVote}(p_i) = \begin{cases}
        a & \text{if } \operatorname{Score}(a) > \operatorname{Score}(b) \\
        b & \text{if } \operatorname{Score}(a) < \operatorname{Score}(b) \\
        0 & \text{otherwise}
        \end{cases}
%     }
% }
\end{equation}
where \texttt{Score}$(a)$ and \texttt{Score}$(b)$ denote the score (judgement) of $a$ and $b$ made by an \xjudgesystem, respectively.

Finally, we calculate the accuracy of an \xjudgesystem following:
\begin{equation}
\label{eqn:acc}
\resizebox{0.9\hsize}{!}{
    \ensuremath{
        \operatorname{Accuracy} = \frac{\sum_{i=1}^n \mathbb{I}[\operatorname{MetricVote}(p_i)=\operatorname{JudgeVote}(p_i)]}{n}
    }
}
\end{equation} 
where $n$ is the total number of pairs in $\mathcal{P}$.

\subsection{Results}

% ?? Any other part I should do? okie, thanks, no problem! What about 'modified' principle assuming there is any? asuming no, I ususally deleted directly instead of modify, ignores minor modification. gotcha :3 thanks! thanks hehe
% fill out the tab:pinciple-diff, just added, 
% the table is to show the usefulness of the general prin and  reudce the effort ofr manaul craft.

%\noindent \textit{Productivity of developing \xjudgesystems.} 

\begin{Summary}{ Productivity of developing \xjudgesystems}{}
Our framework leads to an increased productivity by facilitating the reuse of principles and reducing the development effort. 
\end{Summary}

\vspace{0.1cm}
% \dayi{make it a summary box that directly points out implications, e.g., our framework leads to an increased productivity, by facilitating reuse, reducing development effort, ...  Same for the quality part}

% \dayi{might be worth to add a small note of the developer experience, just essentially say we did an informal interview with our internal pilot user (with x years of FMware development experience. ask sky.) feels that the productivity is improved etc. ...}

\textbf{In the \xjudgesystem with our proposed framework, the majority (i.e., an average of 58\%) of the general principles that are generated in Stage I are reused in the contextualized constitutions (Stage II) across the 5 programming languages.} Table~\ref{tab:principle-diff} details the differences between the principles that are generated in Stage I and those that are revised, manually reviewed and verified in Stage II. Given that Stage I operates automatically, the large numbers of percentages of the reusable principles suggest that the framework can effectively address the challenges B.1 and B.2 (e.g., need reusable requirements and generic criteria) mentioned in Section~\ref{sec:lifecycle} and reduce manual effort. For example, ``\textit{Explain the Why, Not Just the What}'' is one of the reused general principles for writing commit messages across the 5 programming languages. The details of the why and what principle indicate detailed criteria, such as ``\textit{In the body of the commit message, provide `context' by explaining `why' a change was made, not just `what' was done.}'' This reused general principle for CMG is aligned with the finding in prior work~\cite{shen2016automatic, zhang2024automatic}, since without the what and why in commit messages would increase communication cost between developers in a team or community.
% have reported the importance of including the why and what in the commit messages. 
% , resents one criteria of the principle. 
% The result suggests the reduction of manual effort on the generation as 
% Since generating criteria and metrics is time-consuming and labor-intensive, the result suggests the reduction of manual effort on the generation. 
In addition, developers of \xjudgesystems could share a large percentage of principles across the teams building \xjudgesystems for similar types of \xapp, suggesting reducing the manual effort and time of building \xjudgesystems in a company. The high percentages of reusable principles also indicate the high quality of generated principles through the iterations of critiques and revisions in our framework, similar to the prior work~\cite{bai2022constitutional}.

\begin{table}[t]
    \centering
    \renewcommand{\arraystretch}{1.1}
    \caption{The comparison between the general principles and the context-specific principles, i.e., diffs, across the 5 programming languages.}
    \label{tab:principle-diff}
    \begin{tabular}{|l|r|r|r|} 
        \hline
        \multirow{2}{*}{\textbf{Language}} & \multicolumn{3}{c|}{\textbf{\# Principles}} \\ \cline{2-4}
        & \textbf{Reused (\%)$^*$} & \textbf{Added (\%)$^*$} & \textbf{Deleted (\%)$^*$} \\ \hline
        C++          & 10 (59\%) & 4 (24\%) & 3 (18\%)  \\ 
        C\#          & 10 (59\%) & 5 (29\%) & 3 (18\%)  \\ 
        Java         & 9 (53\%) & 3 (18\%) & 5 (29\%) \\ 
        Python       & 9 (53\%) & 5 (29\%) & 5 (29\%)  \\ 
        JavaScript   & 11 (64\%) & 3 (18\%) & 2 (12\%) \\ 
        \hline
    \end{tabular}
    \begin{tablenotes}
        \item[*]$^*$ The percentages indicate the number in a cell over the number of the general principles (i.e., 17).
    \end{tablenotes}

\end{table} 

Additionally, the small numbers of the added and deleted principles in Table~\ref{tab:principle-diff} suggest that developers do not invest considerable effort in obtaining an optimized set of context-specific principles in Stage II, compared to manual generation from scratch (e.g., from 0 to 14 principles). For example, the principle ``\textit{Framework and Version Upgrades}'' was added to the C\# constitution as the \textit{.NET} framework version affects the compatibility and stability of a software project and should be recorded when it has been changed. Similarly, in the Java constitution, the principle ``\textit{Code Style Adherence}'' along with the details ``\textit{Refactor to conform with Google Java Style Guide}'' was added and it suggests that any style changes should be stated and the importance of following best practices in Java. 
% specifically referencing the \textit{.NET} framework. 
% During the revision of Java's constitution, the revisers introduced the principle of \textit{Code Style Adherence}, which states that any style changes should be stated, providing a commit message example such as \textit{Refactor to conform with Google Java Style Guide}. 
In addition, these context-specific principles could be directly transformed into FM-based metrics that the \xjudgesystem advocates for use in rendering judging architectures in Stage III. This transformation suggests the reduction of the labor-intensive and ad-hoc creation of such FM-based metrics for NLP tasks. We did interviews with our internal pilot users. One of our internal users, with approximately 2 years of \xapp development experience, states \textit{``It's pretty good. I reduced the time of collecting information to write the principle. It's `giving' the input if the users want something but don't know exactly what they want.''} 
% the creation of generating such FM-based metrics in a labor-intensive and ad-hoc manner for NLP tasks. 

\vspace{-0.1cm}

%\smallskip \noindent \textit{.}
\begin{Summary}{ Quality of the resulted \xjudgesystems}{}
Our framework leads to an increased accuracy by fine-grained evaluation methods and scoring.
\end{Summary}
\vspace{0.1cm}

\textbf{The accuracy of the judgments made by the \xjudgesystem developed with our proposed framework outperforms those made by the \xjudgesystem developed without our proposed framework, by up to 6.2\%.} Since we pair every data point in each dataset of a particular programming language and obtain approximately 71K relative comparisons to evaluate the accuracy of the judgments, such a large number suggests a robust analysis to ensure the quality of the judgments made by the \xjudgesystem developed with our proposed framework via a systematic verification. In addition, Table~\ref{tab:acc_judgements} indicates the accuracy of the judgments across all 5 programming languages is higher than that of randomly assigning a class in a three-categorie classification (i.e., 33\%).

\vspace{-0.1cm}

\begin{table}[t]
\centering
\renewcommand{\arraystretch}{1.1}
\caption{The \xjudgesystem developed with our proposed framework outperforms the one developed without our proposed framework across the 5 popular programming languages, in terms of accuracy (calculated by Eq.~\ref{eqn:acc}).}
\label{tab:acc_judgements}
\begin{tabular}{|p{1.2cm}|r|r|r|} 
\hline
\multirow{2}{*}{\textbf{Language}} & \multirow{2}{*}{\textbf{\# pairs in $\mathcal{P}$}} &\multicolumn{2}{c|}{\textbf{Accuracy}} \\ \cline{3-4}
 & & \textbf{w/o framework} & \textbf{w/ framework}  \\ \hline
C++ & 70,876 & 37.8  & 43.9 \\ 
C\# & 70,876 & 37.0 &  43.2 \\  
Java & 70,876 & 37.5 & 38.4 \\ 
Python & 71,631 & 41.7 & 46.1 \\  
JavaScript & 71,631 & 41.7 & 45.7 \\  \hline
\end{tabular}
\end{table}

\section{Conclusion}
\label{sec:conclusion}

The goal of this paper is to raise awareness of challenges in the lifecycle of developing \xjudgesystems that are observed in an industrial setting. The proposed framework aims to address major challenges, such as reducing manual effort, and adapts to the fast evolution of FMs and \xapps. We hope the framework enables open discussions and knowledge sharing to advance \xapp evaluation and aligns with the pressing needs for software quality and AI safety from researchers in industry and academia. Given the dynamics of \xapps and their associated FMs and supply chains, developers are imposed to new techniques and challenges. We hope that the disclosure of our challenges and solutions will encourage further explorations and resolutions around \xjudgesystems. We will explore Stages III and IV of our framework and apply them on real-world data. We envision a collaborative effort among developers, researchers, and industry practitioners to refine and expand the framework's capabilities. Such collaboration will promote continuous learning and innovation, thereby enhancing the robustness and scalability of \xjudgesystems.

% \newpage
\footnotesize
\balance
\bibliographystyle{IEEEtranS}

\bibliography{latex/main.bib}

\end{document}